\magnification   \magstep1
\centerline {\bf Renormalization of Wilson Operators in the Light-Cone
                  Gauge }
\vskip 2cm
\centerline   {  A. Andra\v si }
\centerline  {\it "Rudjer Bo\v skovi\' c" Institute, Zagreb, Croatia }
\vskip 4cm

\beginsection Abstract

We test the renormalization of Wilson operators and the Mandelstam-
Leibbrandt gauge in the case when the sides of the loop are parallel
to the $n$, $n*$ vectors used in the M-L gauge. Graphs which in the
Feynman gauge are free of ultra-violet divergences, in the M-L gauge
show double divergences and single divergences with non-local ${\rm Si}$ and
${\rm Ci}$ functions. These non-local functions cancel out when we add
all graphs together and the constraints of gauge invariance are satisfied.
In Appendix C we briefly discuss the problems of the M-L gauge for loops
containing spacelike lines.
\vskip 1cm
 PACS: 11.15.Bt; 11.10.Gh
\vskip 0.5cm  
Keywords: Renormalization; Wilson operators
\vskip 5cm
Electronic address:andrasi@thphys.irb.hr

\vfill \eject

\beginsection 1. Introduction.

The aim of this research is to test the Mandelstam-Leibbrandt gauge,
which is the best form of the lightcone gauge with the condition
$ n\cdot A=0 $, $ n^2=0 $ (where $n $ is a vector used to define the gauge).
\vskip 0.5cm
The Wilson operator is defined as

$$ W = Tr P\exp (-ig\int_C A\cdot dx) \eqno(1) $$
where $ C $ is a closed curve, $ P $ denotes operator and matrix
ordering along $ C $, and the non-abelian gauge field $ A_{\mu}$
is a matrix in some representation $ R $ of the gauge group $ G $.
The path-ordered phase factors (1) are gauge-invariant objects and
therefore an ideal laboratory for testing different gauges.
Also they are better coordinates in a non-abelian theory than are the
conventional vector gauge field matrices $ A_{\mu}(x) $, even though
they are functions of all closed paths. The gauge-variant gauge fields
greatly overdescribe the observable dynamics. The operators (1) are in contrast
gauge-invariant, more precisely describe the dynamics and satisfy
gauge-invariant equations. The hope has therefore arisen that the $ W's $
can replace the $ A's $ as fundamental dynamical variables, and correspondingly
that the loop functions (Wilson operators) can replace the Green's functions.
\vskip 0.5cm
  However, the loop functions are perturbatively even more divergent than the
Green's functions [1]. Therefore, to make any sense out of the above
program, one must renormalize. This has already been done in Lorentz gauges
[2], [3]. In this paper we discuss the renormalization of Wilson operators
in the Mandelstam-Leibbrandt light-cone gauge which became popular with the
revival and intensive research of string theories. The complexity of the
explicit calculations of individual graphs to order $ g^4 $ speaks about
usefulness of the light-cone gauge in perturbative QCD for itself. Apart
from that, in Appendix C, we explain why strict application of dimensional
regularization in the light-cone gauge is not possible in the case of spacelike
and/or timelike lines.
\vskip 0.5cm
It was noted [1] that if the Wilson loop contains a straight light-like
segment, charge renormalization does not work in a simple graph-by-graph way,
but does work when certain graphs are added together. In the M-L gauge
renormalization is even more complicated.   
We shall show to order $ g^4 $ in perturbation theory that $W$
in the M-L gauge obeys multiplicative renormalizability required
 in [1],
$$ W_R(A_R;g_R) = Z(\epsilon)W_B(A_B;g_B,\epsilon) \eqno(2) $$
where suffices $ R $ and $ B $ denote renormalized and bare
quantities and dimensional regularization with $ d=4-\epsilon $
is used. The relationship between $ g_B $ and $ g_R $ and between
$ A_R $ and $ A_B $ should be the same as in ordinary perturbation
theory. $ Z(\epsilon) $ is determined from the vacuum expectation
value $ \langle W \rangle $ 
$$ \langle W_R(g_R)\rangle = Z(\epsilon) \langle W_B(g_B,\epsilon)
 \rangle. \eqno(3) $$
However, the divergences of the individual graphs are not of the
short distance nature and are non-local on the curve $ C $.
They are grouped into four tensors $ n^*_{\beta}n^*_{\rho} $,
$ n_{\beta}n^*_{\rho} $, $ n^*_{\beta}n_{\rho} $ and
$ n_{\beta}n_{\rho} $. There are no transverse divergences of the
type e.g. $ n_{\beta}P_{\rho} $, $ P_{\beta}P_{\rho} $, (we use
the decomposition of the momentum $ p_{\rho} = {1\over 2}n^*_{\rho}
p_{+} + {1\over 2}n_{\rho}p_{-} + P_{\rho} $), neither $ g_{\beta \rho} 
$ divergences, as argued in Appendix A.
\vskip  0.5cm
 Let the divergent part of the amplitude for the emission of two real gluons
 in momentum space be
$$ M_{\beta \rho}=An_{\beta}n_{\rho} +Bn_{\beta}n^*_{\rho}
+Cn^*_{\beta}n_{\rho}+Dn^*_{\beta}n^*_{\rho},  $$  
where $ M $ is the coefficient of the two fields when we expand
$ W $ in terms of the fields and $ n $ and $ n^* $ are the lightlike vectors
used to define the gluon propagator in the M-L gauge 
$$ G_{\beta \rho}=(k^2+i\eta)\{-g_{\beta \rho}+{{n_{\beta}k_{\rho}+
n_{\rho}k_{\beta}}\over{n\cdot k +i\omega n^*\cdot k}}\}.   \eqno (4) $$
The polarization vectors should satisfy
$$ p\cdot e=0 , q\cdot f=0 $$  and can be chosen to satisfy
$$ n\cdot e=n\cdot f=0, $$ 
e.g.
 $$ e_{\beta}={{p_{+}n_{\beta}^*-p_{-}n_{\beta}
 -2p_{\beta}}\over{{|4p_+p_-|}^{1/2}}},
f_{\rho}={{q_{+}n_{\rho}^*-q_{-}n_{\rho}-2q_{\rho}}\over{{|4q_+q_-|}^{1/2}}} \eqno(5) $$
for  $p$ and $q$ on shell respectively. The other independent polarization
vector which is perpendicular to $ P $, $ n $ and $ n^* $ and counterpart to 
$ e $, gives zero when contracted into $ M_{\beta \rho}$ and so plays no role
in this paper. Of course, there is a counterpart of $f$ as well.
\vskip  0.5cm
We have four identities following from gauge invariance, which
the amplitude should satisfy. \hfill\break 
 (a)  $M^{\beta\rho}e_{\beta}f_{\rho} $
    should be the same as Feynman gauge when the external momenta
    $ p $ \break 
\line{\hskip 2.2cm and $ q $ are on shell,\hfill} 
(b) $ M^{\beta\rho}e_{\beta}q_{\rho} =0 $ for $ p $ on shell, \hfill\break
\line{\hskip 0.6cm  $ M^{\beta\rho}p_{\beta}f_{\rho}=0$ for $q$ on shell,\hfill}
(c) $ M^{\beta\rho}p_{\beta}q_{\rho}=0 $.  \hfill (6) \break 
These equations allow us to redefine the vectors (5) into
$$ e'_{\beta}=p_-n_{\beta}-p_+n_{\beta}^*, f'_{\rho}=q_-n_{\rho}
-q_+n_{\rho}^*, $$ and the tensor structure (4) which satisfies
(b) and (c) can be written in the form $$ M_{\beta \rho}=
e'_{\beta}f'_{\rho}M. \eqno(7) $$ 
The form (7) will in section 5 be crucial to show that non-local divergences
must cancel.  
$ (a) $, $(b)$ and $(c)$ impose the constraint on (4)
$$p_+q_+A+p_+q_-B-p_-q_+C-p_-q_-D=0, $$
$$ p_+q_+A-p_+q_-B+q_+p_-C-p_-q_-D=0, $$
$$ p_+q_+A+p_+q_-B+p_-q_+C+p_-q_-D=0, $$ i.e.

$$ q_{+}A + q_{-}B = 0 ,\eqno(8) $$ 
and fix all the ratios of $ A:B:C:D $. The answers we find in eqs.
(15), (22) and (32) confirm this prediction and are invariant under
$ n -> cn $, $n^* ->cn^*$, $p_+ -> cp_+$, $p_- -> cp_-$ for any constant $c$.
\vskip 0.5cm
$A$, $B$, $C$ and $D$ turn out to be local, although from the
example of the self-energy graph [4], one might have expected non-local
divergences to occurr with the Wilson loop. If we take a self-energy
part and try to derive an on-shell physical thing, we get zero
(we take $ S_{\beta \rho}(p)$, put $ p^2 = 0 $, multiply by
$ e_{\beta}e_{\rho} $ where $ e $ is a polarization vector satisfying
$ p\cdot e = 0 $). Therefore we cannot deduce much by arguing that
physical things are gauge invariant - we get just $ 0=0 $.
But, for the Wilson loop, we {\bf do not} get zero if we put
$ p^2=q^2=0 $ and multiply by polarization vectors. So the 
gauge-independence argument does give some information.
As the Feynman gauge non-local divergences cancel [1], $ (a)$,
$(b)$ and $(c)$ explain why there are no non-local divergences in the
M-L Wilson operator.
\vskip  0.5cm
 The abelian $ C_RC_R $ part obeys the factorization
theorem [5], [6]. Therefore in this work we shall concentrate
only on the non-abelian $ C_GC_R $ part of the graphs, where $ C_R $
and $ C_G $ are the Casimirs for the representation used to define
the Wilson loop and the gluons. In the following sections we list final results
for the amplitude $ M_{\beta \rho} $ to order $ g^4 $ after the decomposition
in (4). The graphs are grouped into sets according to their topological
equivalence.

\beginsection  2. The $ n^*_{\beta}n^*_{\rho}$ sector of $ W_B $

We list the final results for groups of graphs shown in the figures.
 The multiplication by the overall factor
$ C_{\beta \rho} $ is understood for each graph. $$ M_{\beta \rho} =
 C_{\beta \rho}M , $$  $$ C_{\beta \rho} = {2\over{\epsilon}}g^4C_G Tr
 (t_b t_d)n^*_{\beta}n^*_{\rho}{\pi}^{2-{{\epsilon}\over 2}}{(2\pi)}^{-n}$$
 We denote the frequent non-local functions which appear in all equations by
$$ {\rm Ci}(x)=\int_{0}^{x}{{\cos t-1}\over t}dt $$ 
$$ {\rm Si}(x)=\int_{0}^{x}{{\sin t}\over t}dt \eqno(9) $$
\line {\bf A-set  \hfill}
\vskip 0.5cm
The graphs contributing to the A-set are shown in Fig.1. There are also
graphs with $p$ and $q$ interchanged.
Then the ultra-violet divergent part of the graphs in Fig.1 is
$$ (M_1+M_2)(A)=-{1\over{p_-q_-}}(e^{-ip_+T}+e^{-iq_+T})(e^{-iq_-L}-1)
\times\{2(e^{-ip_-L}-1) $$ $$ +(e^{-ip_-L}-1){\rm Ci}(p_-L) +i(e^{-ip_-L}+1)
{\rm Si}(p_-L)\} \eqno(10) $$. 
\vskip 0.5cm
\line {\bf B - set  \hfill}
\vskip 0.5cm
$ B $-graphs are shown in Fig.2. Addition of symmetric graphs is understood.
$$ (M_1+M_2+...+M_6)(B)=(e^{-iTr+}+1)\times\{{2\over{p_-q_-}}(e^{-ip_-L}-1)
(e^{-iq_-L}-1) $$ $$ +({2\over{p_-q_-}}+{2\over{p_-r_-}})[(e^{-ir_-L}+1)
{\rm Ci}(r_-L) +i(e^{-ir_-L}-1){\rm Si}(r_-L)] $$ $$ -({2\over{p_-q_-}} +
{2\over{p_-r_-}})[(e^{-ir_-L}+1){\rm Ci}(q_-L)+i(e^{-ir_-L}-1){\rm Si}(q_-L)]
$$ $$-{2\over{p_-r_-}}[(e^{-ir_-L}+1){\rm Ci}(p_-L) +i(e^{-ir_-L}-1){\rm Si}
(p_-L)] $$ $$ -{1\over{p_-q_-}}(e^{-iq_-L}+1)[(e^{-ip_-L}+1){\rm Ci}(p_-L)
+i(e^{-ip_-L}-1){\rm Si}(p_-L)]\} $$ where $$ r=p+q. \eqno(11)  $$
\line {\bf C - set  \hfill}
\vskip 0.5cm
The $ C $ graphs are shown in Fig.3.
$$ (M_1+M_2+M_3+M_4)(C)= {2\over{q_-r_-}}(e^{-iTp_+}+e^{-iTq_+}) $$ $$ \times
\{
(e^{-ir_-L}+1)[{\rm Ci}(q_-L) -{\rm Ci}(p_-L)]+i(e^{-ir_-L}-1)[{\rm Si}(q_-L) -
{\rm Si}(p_-L)]\}  $$ $$ +{2\over{q_-r_-}}(e^{-iTr_+}+1)[(e^{-ir_-L}+1){\rm Ci}
(r_-L) 
+ i(e^{-ir_-L}-1){\rm Si}(r_-L)] $$ $$ -{1\over{q_-p_-}}(e^{-iTp_+}-1)(e^{-iTq_+}
 -1)(e^{-iLq_-}+1)[(e^{-ip_-L}+1){\rm Ci}(p_-L) + i(e^{-ip_-L}-1){\rm Si}(p_-L)]
$$ $$ +{{i\pi}\over{p_-q_-}}(e^{-iTp_+}+1)(e^{-iTq_+}-1)(e^{-ip_-L}-1)
 (e^{-iq_-L}+1) $$ $$ -{{2i\pi}\over{q_-r_-}}(e^{-iTp_+}+1)(e^{-iTq_+}-1)
 (e^{-ir_-L}-1) \eqno(12) $$
Again we have to add the symmetric graphs with $ p $ and $ q $ interchanged. 
\vskip 0.5cm
\line  {\bf  D - set  \hfill}
\vskip 0.5cm
The complete set of $ D $ graphs (including symmetric graphs) is shown in
Fig.4.

$$ (M_1+M_2+... +M_8)(D)= -{2\over{p_-q_-}}(e^{-iTr_+}+1) $$ $$ \times
\{2(e^{-ir_-L}+1){\rm Ci}(r_-L) +2i(e^{-ir_-L}-1){\rm Si}(r_-L) $$ $$
-(e^{-ip_-L}+1)[(e^{-iq_-L}+1){\rm Ci}(q_-L)+i(e^{-iq_-L}-1){\rm Si}(q_-L)]$$ $$
-(e^{-iq_-L}+1)[(e^{-ip_-L}+1){\rm Ci}(p_-L) +i(e^{-ip_-L}-1){\rm Si}(p_-L)]\}$$
$$-2({1\over{q_-r_-}}-{1\over{p_-r_-}})(e^{-iTp_+}+e^{-iTq_+})$$ 
$$ \times \{(e^{-ir_-L}+1)
[{\rm Ci}(q_-L) -{\rm Ci}(p_-L)]+i(e^{-ir_-L}-1)[{\rm Si}(q_-L)-{\rm Si}(p_-L)]
\} $$ $$ -{2\over{p_-q_-}}(e^{-iTp_+}+e^{-iTq_+})[(e^{-iq_-L}+1){\rm Ci}(q_-L)
+i(e^{-iq_-L}-1){\rm Si}(q_-L)] $$ $$ -{2\over{p_-q_-}}(e^{-iTp_+}+e^{-iTq_+})
[(e^{-ip_-L}+1){\rm Ci}(p_-L)+i(e^{-ip_-L}-1){\rm Si}(p_-L)] $$ $$ - 
{2\over{p_-q_-}}(e^{-iTp_+}+e^{-iTq_+})(e^{-ip_-L}-1)[{\rm Ci}(q_-L)-
i{\rm Si}(q_-L)] $$ $$ -{2\over{p_-q_-}}(e^{-iTp_+}+e^{-iTq_+})(e^{-iq_-L}-1)
[{\rm Ci}(p_-L) -i{\rm Si}(p_-L)] $$ $$ -{{2i\pi}\over{p_-q_-}}(e^{-iTp_+}
-e^{-iTq_+})(e^{-iq_-L}-e^{-ip_-L}) $$ $$
-2i\pi({1\over{p_-r_-}}-{1\over{q_-r_-}})(e^{-ir_-L}-1)(e^{-iTq_+}-
e^{-iTp_+}) \eqno(13) $$ 
\line {\bf  E - set   \hfill}
\vskip  0.5cm
$ E $-set is shown in Fig.5.
$$ (M_1+...+M_4)(E)= -4(e^{-iTr_+}+1)\times \{{1\over{p_-q_-}}
[(e^{-ir_-L}+1){\rm Ci}(r_-L)+i(e^{-ir_-L}-1){\rm Si}(r_-L)] $$  $$
-{1\over{q_-r_-}}[(e^{-ir_-L}+1){\rm Ci}(p_-L)+i(e^{-ir_-L}-1){\rm Si}(p_-L)]
$$  $$ -{1\over{p_-r_-}}[(e^{-ir_-L}+1){\rm Ci}(q_-L)+i(e^{-ir_-L}-1)
{\rm Si}(q_-L)]\}  \eqno(14)  $$
The complete sum of all the graphs contributing to the $ n^*n^* $ sector is very
simple.
$$ S_{\beta \rho}(n^*n^*) = {8\over{\epsilon}}g^4C_GTr(t_b t_d)n_{\beta}^*
n_{\rho}^*{\pi}^{2-{{\epsilon}\over 2}}{(2\pi)}^{-4} $$ $$ \times
{1\over{p_-q_-}}(e^{-iTq_+}-1)(e^{-iTp_+}-1)(e^{-iq_-L}-1)(e^{-ip_-L}-1)
$$ $$ ={8\over{\epsilon}}g^2C_G\pi^2{(2\pi)}^{-4}B_{\beta \rho}(n^*n^*)   
 \eqno(15)  $$ where $ B_{\beta \rho}(n^*n^*) $ denotes the Born term which is
 in the $ n^*n^* $ sector only.
The non-local functions have cancelled out. This result alone does not prove
renormalizability as the field renormalization matrix in the lightcone gauge
mixes all three sectors.

\beginsection   3. The $ n_{\beta}n_{\rho} $ sector of $ W_B $

Again we list the final results, but here the overall factor is
$$ C'_{\beta \rho}={2\over{\epsilon}}g^4C_G Tr(t_bt_d)n_{\beta}n_{\rho}
{\pi}^{2-{{\epsilon}\over 2}}{(2\pi)}^{-n}{1\over{q_+p_+}}, $$ $$
M_{\beta \rho}= C'_{\beta \rho}M  \eqno(16)  $$
\line  {\bf  G2 - set  \hfill}
\vskip  0.5cm
The sum of the three graphs of the G2-set shown in Fig.(6) is
$$ M(G2)= -{8\over{\epsilon}}(e^{-ir_-L}+1)(e^{-iTp_+}-1)(e^{-iTq_+}-1) $$
$$ -2(e^{-iTp_+}-1)(e^{-iTq_+}-1)\{(e^{-ir_-L}+1)[{\rm Ci}(r_-L) +
2\ln (TL{\mu}^2) +i\pi]$$ $$ +i(e^{-ir_-L}-1){\rm Si}(r_-L)\} $$  $$
-2(e^{-iTq_+} + e^{-iTp_+})\{(e^{-ir_-L}+1)[{\rm Ci}(r_-L)-{\rm Ci}(q_-L)
-{\rm Ci}(p_-L)] $$ $$+i(e^{-ir_-L}-1)[{\rm Si}(r_-L)-{\rm Si}(q_-L)
-{\rm Si}(p_-L)]\}  \eqno(17)  $$
\line  {\bf  G1 - set  \hfill}
\vskip  0.5cm
G1-set of graphs are the graphs with one 3-gluon vertex. There are two
groups of such graphs. The two graphs shown in Fig.7 give
$$  M^a(G1)=(e^{-iTp_+}-1)(e^{-iTq_+}-1)(e^{-iq_-L}+1) $$ $$ \times
\{(e^{-ip_-L}+1)[{\rm Ci}(p_-L)+2\gamma +{4\over{\epsilon}}+i\pi
+2\ln(TL{\mu}^2)] +i(e^{-ip_-L}-1){\rm Si}(p_-L)\}. \eqno(18)  $$
Of course the graphs with $ p $ and $ q $ interchanged must be added.
The graphs in Fig.8 give
$$ M^b(G1)=(e^{-iTp_+}-1)(e^{-iTq_+}-1)(e^{-iq_-L}-1) $$  $$ \times
\{2(e^{-ip_-L}-1) + (e^{-ip_-L}-1){\rm Ci}(p_-L) + i(e^{-ip_-L}+1){\rm Si}
(p_-L)\}. \eqno(19)  $$
Again there is a symmetric set of graphs with $ p $ and $ q $ interchanged.    
\vskip  0.5cm
\line  {\bf  G(L+R) - set  \hfill}
\vskip  0.5cm
Adding the symmetric graphs to Fig.9, the total sum of 4 graphs is
$$ G(L+R)=-2(e^{-iTr_+}+1)\{(e^{-ir_-L}+1)[{\rm Ci}(p_-L)+{\rm Ci}(q_-L)
-{\rm Ci}(r_-L)] $$  $$ +i(e^{-ir_-L}-1)[{\rm Si}(p_-L)+{\rm Si}(q_-L)
-{\rm Si}(r_-L)]\}.  \eqno(20)  $$

\line  {\bf  G0 - set  \hfill}
\vskip  0.5cm
Fig.10 gives
$$  M(G0)=-{8\over{\epsilon}}(e^{-iTp_+}-1)(e^{-iTq_+}-1)(e^{-iq_-L}+
e^{-ip_-L}) $$  $$\times {({\mu}^2TL)}^{{\epsilon}\over 2}(1+\gamma 
{{\epsilon}\over 2}+{{i\pi \epsilon}\over 4}).  \eqno(21)  $$
The sum of all the graphs contributing to the $ n n $ sector is
$$ S_{\beta \rho}(nn)={8\over{\epsilon}}(e^{-iTp_+}-1)(e^{-iTq_+}-1)
(e^{-ip_-L}-1)(e^{-iq_-L}-1){1\over{q_+p_+}}  $$  $$ \times g^4C_GTr(t_bt_d)
n_{\beta}n_{\rho}{\pi}^{2-{{\epsilon}\over 2}}{(2\pi)}^{-4}$$
$$={8\over{\epsilon}}g^2C_G\pi ^2{(2\pi)}^{-4}B_{\beta \rho}(nn)
 \eqno(22) $$ where $B_{\beta \rho}(nn) $ is the $ g^2 $ term for the
$nn$ sector of $ W_B(g_B,\epsilon) $.

\beginsection  4. The $ n_{\beta}n^*_{\rho} $ sector of $ W_B $

The final results for the $ n_{\rho}n^*_{\beta} $ sector we get from $
n_{\beta}n^*_{\rho} $ by the change $ p, b, \beta $ into $ q, d, \rho. $
The overall factor for all the graphs in this sector is
$$ C_{\beta \rho}"= {2\over{\epsilon}}g^4C_GTr(t_bt_d)n_{\beta}n^*_{\rho}
{\pi}^{2-{{\epsilon}\over 2}}{(2\pi)}^{-n}, $$  $$ M_{\beta \rho}=
C_{\beta \rho}"M.  \eqno(23) $$

\line  {\bf  A - set  \hfill}
\vskip 0.5cm
The graphs in Fig.11 give
$$ M(A)=(e^{-iTq_+}+e^{-iTp_+})(e^{-iq_-L}-1){1\over{p_+q_-}} $$  $$
\times \{2(e^{-ip_-L}-1)+(e^{-ip_-L}-1){\rm Ci}(p_-L) + i(e^{-ip_-L}+1)
{\rm Si}(p_-L)\}. \eqno(24)  $$

\line  {\bf  A' - set  \hfill}
\vskip 0.5cm
The $ A' $-set is presented in Fig.12.
$$ M(A')=-(e^{-iTp_+}-1)(e^{-iTq_+}-1)(e^{-ip_-L}-1){1\over{p_+q_-}}$$ $$ \times
\{2(e^{-iq_-L}-1)+(e^{-iq_-L}-1){\rm Ci}(q_-L)+i(e^{-iq_-L}+1){\rm Si}(q_-L)
\eqno(25)  $$

\line  {\bf  B - set  \hfill}
\vskip  0.5cm
$ B $- set is shown in Fig.13.
$$ M(B)=-(e^{-iTr_+}+1)\times \{{2\over{p_+q_-}}(e^{-ip_-L}-1)(e^{-iq_-L}-1)
$$  $$ +({2\over{p_+q_-}}+{2\over{p_+r_-}})[(e^{-ir_-L}+1){\rm Ci}(r_-L)
+i(e^{-ir_-L}-1){\rm Si}(r_-L)] $$ $$-({2\over{p_+q_-}}+{2\over{p_+r_-}})
[(e^{-ir_-L}+1){\rm Ci}(q_-L)+i(e^{-ir_-L}-1){\rm Si}(q_-L)]  $$ $$
+{2\over{p_+r_-}}[(e^{-ir_-L}+1){\rm Ci}(p_-L)+i(e^{-ir_-L}-1){\rm Si}(p_-L)]
$$ $$ -{1\over{p_+q_-}}(e^{-iq_-L}+1)[(e^{-ip_-L}+1){\rm Ci}(p_-L)
+i(e^{-ip_-L}-1){\rm Si}(p_-L)]\}  \eqno(26)  $$

\line {\bf  C - set   \hfill}
\vskip 0.5cm
Graphs grouped into the $ C $-set are shown in Fig.14.
$$ M(C)={2\over{p_+q_-}}[{2\over{\epsilon}}+\ln (TL{\mu}^2)+\gamma]
(e^{-iTp_+}-1)(e^{-iTq_+}-1)(e^{-ip_-L}-1)(e^{-iq_-L}-1) $$ 
$$ +2(e^{-iTr_+}+1)({1\over{q_-p_+}}+{1\over{r_-p_+}})\{(e^{-ir_-L}+1){\rm Ci}
(r_-L)+i(e^{-ir_-L}-1){\rm Si}(r_-L)\}$$ $$ -2(e^{-iTp_+}+e^{-iTq_+})({1\over{
q_-p_+}}+{1\over{r_-p_+}})\{(e^{-ir_-L}+1){\rm Ci}(q_-L)+i(e^{-ir_-L}-1)
{\rm Si}(q_-L)\} $$ $$
+\{{2\over{r_-p_+}}(e^{-iTp_+}+e^{-iTq_+})-{1\over{q_-p_+}}(e^{-iTr_+}+1)
(e^{-iq_-L}+1)$$ $$+{1\over{q_-p_+}}(e^{-iTp_+}+e^{-iTq_+})(e^{-iq_-L}-1)\} $$
$$\times\{{\rm Ci}(p_-L)-i{\rm Si}(p_-L)\}$$ $$ +\{{2\over{r_-p_+}}e^{-ir_-L}
(e^{-iTp_+}+e^{-iTq_+})-{1\over{q_-p_+}}e^{-ip_-L}(e^{-iq_-L}+1)(e^{-iTr_+}+1)$$
$$-{1\over{q_-p_+}}e^{-ip_-L}(e^{-iq_-L}-1)(e^{-iTp_+}+e^{-iTq_+})\} $$
$$\times\{{\rm Ci}(p_-L)+i{\rm Si}(p_-L)\} $$
$$+{{i\pi}\over{q_-p_+}}(e^{-iTp_+}-1)(e^{-iTq_+}-1)(e^{-ip_-L}-1)(e^{-iq_-L}
-1) $$ $$ -{{2i\pi}\over{r_-p_+}}(e^{-iTp_+}-1)(e^{-iTq_+}+1)(e^{-ir_-L}-1).
\eqno(27) $$

\line {\bf  C'- set  \hfill}
\vskip  0.5cm
The two graphs of $ C' $-set in Fig.15 give
$$ M(C')={1\over{q_-p_+}}(e^{-iTp_+}-1)(e^{-ip_-L}+1) $$ $$ \times
\{(e^{-iTq_+}-1)[(e^{-iq_-L}+1){\rm Ci}(q_-L)+i(e^{-iq_-L}-1){\rm Si}(q_-L)]
-3i\pi(e^{-iq_-L}-1)\} \eqno(28) $$
\vskip 0.5cm

\line  {\bf  D - set \hfill}
\vskip 0.5cm
$ D $-set is shown in Fig.16.
$$ M(D)=-{2\over{q_-p_+}}[{2\over{\epsilon}}+\ln
(TL{\mu}^2)+{{i\pi}\over2}+\gamma]
\times (e^{-iTp_+}-1)(e^{-iTq_+}-1)(e^{-ip_-L}-1)(e^{-iq_-L}-1) $$
$$-{2\over{q_-p_+}}(e^{-iTp_+}-1)(e^{-iTq_+}-1)(e^{-ip_-L}+1) $$ $$
\times \{(e^{-iq_-L}+1){\rm Ci}(q_-L)+i(e^{-iq_-L}-1)({\rm
Si}(q_-L)-{{\pi}\over2})\} $$
$$ +{{2i\pi}\over{q_-p_+}}(e^{-iTp_+}-1)(e^{-ip_-L}+1)(e^{-iq_-L}-1)  \eqno(29)
$$

\line  {\bf  E - set   \hfill}
\vskip 0.5cm
$ E $-set contains graphs with two 3-gluon vertices and the graph with
 the 4-gluon vertex spanning across the loop. They are shown in Fig.17.
$$ M(E)=-{2\over{p_+r_-}}\times \{(e^{-iTr_+}+1)[(e^{-ir_-L}+1){\rm Ci}(r_-L)
+i(e^{-ir_-L}-1){\rm Si}(r_-L)] $$  $$ +(e^{-iTp_+}+e^{-iTq_+})[(e^{-ir_-L}+1)
({\rm Ci}(p_-L)-{\rm Ci}(q_-L))+i(e^{-ir_-L}-1)({\rm Si}(p_-L)-{\rm Si}(q_-L))]
$$  $$ -i\pi(e^{-ir_-L}-1)(e^{-iTq_+}+1)(e^{-iTp_+}-1)\}  \eqno(30) $$

\line  {\bf  F - set  \hfill}
\vskip 0.5cm
The four graphs of the $ F $-set are shown in Fig.18. Of course, there are
also symmetric graphs with $p$ and $q$ interchanged. The complete sum of eight
graphs amounts to
$$ M(F)={2\over{p_+r_-}}(e^{-iTr_+}+1)\times \{(e^{-ir_-L}+1)[{\rm Ci}(p_-L)
-{\rm Ci }(q_-L)+{\rm Ci}(r_-L)]$$ $$+i(e^{-ir_-L}-1)[{\rm Si}(p_-L)-{\rm
Si}(q_-L)+{\rm Si}(r_-L)]\}. \eqno(31) $$
The total sum for the $ nn^* $ sector is
$$ S_{\beta \rho}(nn^*)=-{8\over{\epsilon}}g^4C_GTr(t_bt_d)n_{\beta}n_{\rho}^*
{\pi}^{2-{{\epsilon}\over 2}}{(2\pi)}^{-n} $$ $$ \times{1\over{p_+q_-}}
(e^{-iTp_+}-1)(e^{-iTq_+}-1)(e^{-ip_-L}-1)(e^{-iq_-L}-1) $$ $$
-{8\over{\epsilon}}g^4C_GTr(t_bt_d)n_{\rho}n_{\beta}^*{\pi}^{2-{{\epsilon}
\over 2}}{(2\pi)}^{-n} $$ $$ \times{1\over{q_+p_-}}(e^{-iTp_+}-1)(e^{-iTq_+}-1)
(e^{-ip_-L}-1)(e^{-iq_-L}-1)  \eqno(32) $$
This is again proportional to the $ g^2 $ term with the same factor
$ {8\over{\epsilon}} $ as in eq.(15) and eq.(22).

\beginsection   5. Discussion

We are now going to explain how (2) works out to order $ g^4 $. The field
renormalization matrix in the M-L gauge is [8],[9],[10],in momentum
space, where $ A(p) $ is the gluon field in momentum space 
$$ A^B_{\beta}(p)=(1+{11\over 6}c)[g _{\beta \gamma}-cn_{\beta}(n^*_{\gamma}-
{{n^*\cdot p}\over{n\cdot p +i\eta n^*\cdot p}}n_{\gamma})]A^{\gamma R}(p) $$ $$
=z_{\beta \gamma}A^{\gamma R}(p)  $$
where  $$ c={{g^2_R}\over{8{\pi}^2\epsilon}}C_G $$ and coupling constant
renormalization is
$$ g_B=(1-{{11}\over 6}c)g_R.  \eqno(33) $$
\vskip 0.5cm
On the right hand side (2) contains various sorts of fourth order terms.
\vskip 0.5cm
\line { (a)  $ W_B $ to fourth order, $ Z, z$ and $g_B$ to zeroth 
order  \hfill}

\line { (b)  $W_B $ to second order, $Z$ to second order  \hfill}

\line { (c)  $ W_B $ to second, $ g_B $ to second  \hfill}

\line { (d)  $W_B $ to second, $ z $ to second  \hfill}

Then (b) contributes only to the abelian $ C_RC_R $ part, while  (c) and
(d) should give the counter-terms needed to cancel the UV divergences we
found in (a). Of course, since $ W_B $ to second order has two real gluons,
that is two $ A_B $ operators, it gets two $ z $ factors, one depending on
$ p $ and the other on $ q $. We list counter-terms for each sector separately. 

\vskip 0.5cm
\line  { (1)  $ n^*n^*$  sector    \hfill}
\vskip 0.5cm  
Although the Born term is contained in the $ n^*n^* $ sector only, we
had to study the off-shell sectors as well. The reason is the field
renormalization matrix $ z_{\beta \gamma} $ which mixes all three sectors.
The Born term to order $ g^2 $ is
$$ B_{\beta
\rho}(n^*n^*)={1\over{p_-q_-}}g_B^2n^*_{\beta}n^*_{\rho}HA_B^{\beta}(p)
A_B^{\rho}(q) $$ where $$ H=Tr(t_bt_d)(e^{-iTp_+}-1)(e^{-iTq_+}-1)
(e^{-iLp_-}-1)(e^{-iLq_-}-1). \eqno(34) $$ To order $ g_R^4 $ the operator
$ (z-1)W_B+(g_B-g_R)W_B $ on the right of eq.(2) gives for the $n^*n^*$
sector the counter-term 
$$ W_{\beta \rho}^{ct}(n^*n^*)=n_{\beta}^*n_{\rho}^*{{g_R^2}\over{p_-q_-}}
H[{{11}\over 6}cg^{\beta \gamma}-cn^{\beta}(n^{\gamma *}-{{n^*\cdot p}\over
{n\cdot p}}n^{\gamma})]A^R_{\gamma}(p)A^R_{\rho}(q) $$ $$
+n_{\beta}^*n_{\rho}^*{{g_R^2}\over{p_-q_-}}H[{{11}\over 6}cg^{\rho \gamma}
-cn^{\rho}(n^{\gamma *}-{{n^*\cdot q}\over{n\cdot q}}n^{\gamma})]
A^R_{\gamma}(q)A^R_{\beta}(p) $$ $$-2n_{\beta}^*n_{\rho}^*{{11}\over 6}
cH{{g_R^2}\over{p_-q_-}}A^{\beta R}(p)A^{\rho R}(q). \eqno(35) $$
We notice that the factor $ {{11}\over 6}c $ cancels out between the wave
function renormalization (two first terms) and the coupling constant
renormalization (last term). Hence, the counter-term to order $ g_R^4 $
for the $n^*n^*$ sector is
$$W_{\beta \rho}^{ct}(n^*n^*)= -4cn^*_{\rho}n^*_{\beta}{1\over{p_-q_-}}H +2c[n^*_{\beta}n{\rho}
{1\over{p_-q_+}}+n^*_{\rho}n_{\beta}{1\over{q_-p_+}}]H. \eqno(36) $$
\vskip 0.5cm
\line {  (2)  $nn^*$ sector  \hfill}
$$W_{\beta \rho}^{ct}(nn^*)= -4cn_{\rho}n_{\beta}{1\over{p_+q_+}}H+2c[n^*_{\beta}n_{\rho}{1\over{p_-q_+}}
+n^*_{\rho}n_{\beta}{1\over{p_+q_-}}]H \eqno(37) $$
\vskip 0.5cm
\line  { (3)  $ nn $ sector gives zero  \hfill}

 The sum of (1) and (2) gives exactly the counter-terms needed to cancel
 (15), (22) and (32). The complications with non-local ${\rm Si}$ and 
 $ {\rm Ci} $
 divergences were caused by the choice of the M-L gauge, not the lightlike
 sides of the Wilson loop, as shown in Appendix C. We certainly expected (a) of
eq.(6) to be gauge-invariant, but in fact we find that the {\bf whole}
of the divergent part of $ M_{\beta \rho} $ is gauge invariant.

\vskip 0.5cm
We shall now explain how the field renormalization
matrix $ z_{\beta \gamma} $ leaves this tensor structure unchanged. Let us
envoke the tensor structure for the amplitude in eq.(7)
$$ M_{\beta \rho}=e_{\beta}'f_{\rho}'M. \eqno(38) $$
This vanishes when contracted with $ p_{\beta} $ or $ q_{\rho}. $
Then it is easy to see why $ z_{\beta \gamma} $ does not change the structure.
The non-local structure in $ z_{\beta \gamma} $ contains
$$ n_{\beta}(n_{\gamma}^*-{{n^*\cdot p}\over{n\cdot p}}n_{\gamma})
=-{{n_{\beta}e'_{\gamma}}\over{p_+}}. \eqno(39) $$
When contracted with $ M_{\beta \rho} $ the term $ {{n\cdot e'}\over{p_+}}=-2$
becomes free of non-localities.
\vskip 0.5cm
 Although we have demonstrated multiplicative
renormalizability of Wilson operators to order $ g^4 $ in the M-L gauge,
the complexity of the actual calculation raises the question of usefulness
of both, lightcone gauge and Wilson operators as fundamental variables in
perturbative QCD. Lightcone gauge has additional problems for loops containing
spacelike and/or timelike lines as explained in Appendix C.

\beginsection  Acknowledgment

  A.A. wishes to thank Prof. J.C. Taylor for the invaluable help and advice
  which made this work possible. The author is grateful to The Royal Society
  for financial help and D.A.M.T.P. for hospitality.
  This work was supported by the Ministry of Science and Technology of the
  Republic of Croatia under Contract No. 00980103.

\beginsection  Appendix A

 There are no transverse components in the Wilson operator as we have assumed
 in (4).
 Let us take one of the characteristic integrals which appears in
the graph with one 3-gluon vertex in Fig.7.
$$ Z_{\beta}=\int d^nk{{2K_{\beta}}\over{k^2{(p-k)}^2k_+}}(e^{-iT(p-k)_+}
-e^{-iTk_+}) $$ $$ \times{1\over{(p-k)_-}}(e^{-ik_-L}-1)(1-e^{-i(p-k)_-L})
=P_{\beta}\times M  \eqno(A1) $$
We multiply both sides by the perpendicular momentum $ P_{\beta} $, and write
$$ 2P\cdot K=K^2-k_+k_--(P-K)^2+(p-k)_+(p-k)_-+p_+k_-+p_-k_+ \eqno(A2) $$

$$ Z\cdot P=P^2\times M=\int d^nk{{(p-k)^2-k^2+p_-k_++p_+k_-}\over
{k^2(p-k)^2k_+}} $$ $$ \times
\{e^{-iTp_+}(e^{iTk_+}-1)-(e^{-iTk_+}-1)+(e^{-iTp_+}-1)\} $$ $$
\times {1\over{(p-k)_-}}\{e^{-ip_-L}(e^{i(p-k)_-L}-1)+e^{-i(p-k)_-L}-1\}
\eqno(A3) $$
\vskip 0.5cm
In this form we can integrate each of the terms in (A3).
$ k_+p_- $ gives UV finite term as the integrals of the type
$$ I=\int d^nk{1\over{k^2(p-k)^2}}e^{iTk_+}{1\over{(p-k)_-}}
(e^{-i(p-k)_-L}-1) \eqno(A4)  $$
contain the oscillating factor $ e^{iTk_+} $ which suppresses the possible 
UV divergences.
Let us denote by $ Y $ the contribution from $ k_-p_+ $.
$$ Y=\int d^nk{{p_+k_-}\over{k^2(p-k)^2k_+}}\times \{e^{-iTp_+}(e^{iTk_+}-1)
-(e^{-iTk_+}-1)+(e^{-iTp_+}-1)\} $$ $$ \times {1\over{(p-k)_-}}\{
e^{-ip_-L}(e^{i(p-k)_-L}-1)+e^{-i(p-k)_-L}-1\}  \eqno(A5)  $$
The factor $ (e^{-iTp_+}-1) $ gives only UV finite term. Also we can write

$$ {{k_-}\over{(p-k)_-}} = {{(k-p)_-+p_-}\over{(p-k)_-}}=-1 \eqno(A6) $$
modulo UV finite terms.    
We change the variable $ p-k=k' $ and use the argument analogous to (A4)
but now with $ k_+ $ and $ k_- $ interchanged. The integral
$$ A=\int d^nk{1\over{k^2(p-k)^2}}e^{ik_-L}{1\over{(p-k)_+}}(e^{-i(p-k)_+T}-1)
\eqno(A7) $$ is UV finite due to the oscillating factor $ e^{ik_-L} $.
\vskip 0.5cm
Therefore the UV divergent part of $ Y $ is
$$ Y=p_+(e^{-ip_-L}+1)\int d^nk{1\over{k^2(p-k)^2(p-k)_+}} $$ $$ \times
\{e^{-iTp_+}(e^{iT(p-k)_+}-1)-(e^{-iT(p-k)_+}-1)\}. \eqno(A8) $$
After the integration over $ k_- $ using the formula
$$ \int d^nk{1\over{k^2(p-k)^2}}=i{\pi}^{2-{{\epsilon}\over 2}}\Gamma
({{\epsilon}\over 2}){(-p^2-i\eta)}^{-{{\epsilon}\over 2}}\int_{0}^{1}dx
x^{-{{\epsilon}\over 2}}{(1-x)}^{-{{\epsilon}\over 2}} $$ where
$$ k_+=p_+x  \eqno(A9) $$ we obtain
$$ Y=i{\pi}^{2-{{\epsilon}\over 2}}\Gamma({{\epsilon}\over 2}){(-p^2-i\eta)}^
{-{{\epsilon}\over 2}}(e^{-ip_-L}+1)\times [(e^{-iTp_+}-1){\rm Ci}(p_+T)
+i(e^{-iTp_+}+1){\rm Si}(p_+T)]. \eqno(A10) $$
\vskip 0.5cm
The remaining two integrals in (A3) we denote by $ E $ and $ F $.
$$ E=\int d^nk{1\over{k^2k_+}}\{ e^{-iTp_+}(e^{iTk_+}-1)-(e^{-iTk_+}-1)
+(e^{-iTp_+}-1)\} $$ $$ \times {1\over{(p-k)_-}}\{e^{-i(p-k)_-L}-1+e^{-ip_-L}(e^{i(p-k)_-L}-1)
\}.  \eqno(A11) $$
The term $ (e^{-iTp_+}-1) $ gives vanishing contribution upon the integration
in the complex $ k_+ $ plane as
$$ \int d^nk{1\over{k^2k_+}}f(p_-,k_-)= 0 \eqno(A12) $$
because both poles lie in the same half-plane with $ k_+ $ regulated in the
sense of Mandelstam [7]. Other terms have no pole at $ k_+=0 $. For the first
$ (e^{iTk_+}-1) $ we close the contour in the upper half-plane and pick up
a pole at $ k_+={{K^2-i\eta}\over{k_-}}\theta (-k_-) $, while for the second
$ (e^{-iTk_+}-1) $ we close the contour in the lower half-plane and pick up
a pole at $ k_+={{K^2-i\eta}\over{k_-}}\theta(k_-) $.
$$ E=i\pi e^{-iTp_+}\int_{-\infty}^{0}dk_- \int d^{2-\epsilon}K
{1\over{K^2-i\eta}}(e^{iT{{K^2-i\eta}\over{k_-}}}-1) $$ $$ \times
{1\over{(p-k)_-}}\{e^{-i(p-k)_-L}-1+e^{-ip_-L}(e^{i(p-k)_-L}-1)\} $$ $$
-i\pi \int_{0}^{\infty}dk_- \int d^{2-\epsilon}K{1\over{K^2-i\eta}}
(1-e^{-iT{{K^2-i\eta}\over{k_-}}}) $$ $$ \times {1\over{(p-k)_-}}\{
e^{-i(p-k)_-L}-1+e^{-ip_-L}(e^{i(p-k)_-L}-1)\} \eqno(A13) $$
In the case of the lightlike Wilson loop we can omit the tadpoles in $ K^2 $ 
of the form
$$ \int d^{2-\epsilon}K{1\over{K^2-i\eta}}f(k_-,p_-) = 0. \eqno(A14) $$
This step is not permitted for the spacelike or timelike lines
(we explain why in Appendix C). Using the integral
$$ T=\int d^{2-\epsilon}K{1\over{K^2-i\eta}}e^{-iT{{K^2-i\eta}\over{k_-}}}
=-{2\over{\epsilon}}{\pi}^{1-{{\epsilon}\over 2}}{({T\over{k_-}})}^
{{\epsilon}\over 2}e^{{i\pi \epsilon}\over 4} \eqno(A15) $$
and evaluating the remaining $ k_- $ integrals, we obtain
$$ E={2\over{\epsilon}}i{\pi}^{2-{{\epsilon}\over 2}}e^{{i\pi \epsilon}\over 4}
T^{{\epsilon}\over 2}{p_-}^{-{{\epsilon}\over 2}}\times \{{2\over{\epsilon}}
(e^{-iTp_+}-1)(e^{-ip_-L}+1) $$ $$ +(e^{-iTp_+}-1)[(e^{-ip_-L}+1){\rm ci}
(p_-L)+i(e^{-ip_-L}-1){\rm si}(p_-L)]-i\pi(e^{-ip_-L}-1) \} \eqno(A16) $$
\vskip 0.5cm
The last integral is
$$ F=\int d^nk{1\over{(p-k)^2k_+}}(e^{-iTk_+}-e^{-iT(p-k)_+}) 
 \times {1\over{(p-k)_-}}(e^{-ik_-L}-1)(1-e^{-i(p-k)_-L}). \eqno(A17) $$
Using the same methods as for the integral $ E $, but here the auxiliary
formula is
$$ \int d^{2-\epsilon}K{1\over{p_++{{K^2-i\eta}\over{k_-}}}}
e^{-iT(p_++{{K^2-i\eta}\over{k_-}})} $$ $$ =-{\pi}^{1-{{\epsilon}\over2}}
{k_-}^{1-{{\epsilon}\over 2}}{\rm Ei}(-iTp_+)=-{\pi}^{1-{{\epsilon}\over 2}}
{k_-}^{1-{{\epsilon}\over 2}}[{\rm ci}(p_+T)-i{\rm si}(p_+T)] \eqno(A18) $$
we get
$$ F=-i{\pi}^{2-{{\epsilon}\over 2}}\Gamma({{\epsilon}\over 2})
{(-p_+p_--i\eta)}^{-{{\epsilon}\over 2}}(e^{-iTp_+}-1)$$ $$\times \{(e^{-ip_-L}+1)
{\rm Ci}(p_-L)+i(e^{-ip_-L}-1){\rm Si}(p_-L)\} $$ $$ +i{\pi}^{2-{{\epsilon}
\over 2}}\Gamma(-{{\epsilon}\over 2})L^{{{\epsilon}\over 2}}(e^{-ip_-L}+1)
\times \{(e^{-iTp_+}-1)[{\rm ci}(p_+T)+{2\over{\epsilon}}{p_+}^{-{{\epsilon}
\over 2}}-\gamma]$$ $$+i(e^{-iTp_+}+1){\rm si}(p_+T)+i\pi e^{-iTp_+}\} $$ 
$$ +i{\pi}^{2-{{\epsilon}\over 2}}L^{{{\epsilon}\over 2}}{p_+}^{-
{{\epsilon}\over 2}}(e^{-iTp_+}+1)(e^{-ip_-L}-1)\times {{i\pi}\over{\epsilon}}.
\eqno(A19) $$
The sum of the pole parts in $ Z\cdot P $ is
$$ Z\cdot P=Y+(E+F)=0, \eqno(A20) $$
hence there are no UV divergences in the transverse momentum $ P_{\beta} $.

\beginsection   Appendix B

As an example of the complications caused by using the M-L gauge, let us
take the diagram shown in Fig.19 which in the Feynman gauge contains no
ultra-violet divergences. In the M-L gauge the UV divergent part of this
graph is
$$ G_{\beta \rho}=-2g^4C_GTr(t_bt_d)n_{\beta}n_{\rho}{\pi}^{2-{{\epsilon}\over
2}}{(2\pi)}^{-n}\times \{{8\over{{\epsilon}^2}}{1\over{q_+p_+}}
e^{-iTq_+}(e^{-iTp_+}-1)(e^{-ir_-L}+1) $$ $$ -{8\over{{\epsilon}^2}}
{1\over{q_+r_+}}(e^{-iTr_+}-1)(e^{-ir_-L}+1) $$
$$+{2\over{\epsilon}}{1\over{q_+p_+}}e^{-iTq_+}(e^{-iTp_+}-1)[(e^{-ir_-L}+1)
({\rm Ci}(r_-L)+2\ln (TL{\mu}^2)+i\pi +2\gamma)$$ $$+i(e^{-ir_-L}-1){\rm Si}
(r_-L)] $$ $$ -{2\over{\epsilon}}{1\over{q_+r_+}}(e^{-iTr_+}-1)[(e^{-ir_-L}+1)
({\rm Ci}(r_-L)+2\ln(TL{\mu}^2)+i\pi +2\gamma)+i(e^{-ir_-L}-1)
{\rm Si}(r_-L)] $$
$$ +{2\over{\epsilon}}{1\over{q_+p_+}}e^{-iTq_+}[(e^{-ir_-L}+1)({\rm Ci}(r_-L)
-{\rm Ci}(q_-L)-{\rm Ci}(p_-L)) $$ $$ +i(e^{-ir_-L}-1)({\rm Si}(r_-L)
-{\rm Si}(q_-L)-{\rm Si}(p_-L))] $$
$$ -{4\over{\epsilon}}{1\over{{r_+}^2}}(e^{-ir_-L}+1)[(e^{-iTr_+}+1)
{\rm Ci}(r_-L)+i(e^{-iTr_+}-1){\rm Si}(r_-L)]\} $$
$$-{8\over{\epsilon}}{1\over{r_+r_-}}g _{\rho \beta}g^4C_GTr(t_bt_d)
{\pi}^{2-{{\epsilon}\over 2}}{(2\pi)}^{-n}\times \{(e^{-iTr_+}+1)[(
e^{-ir_-L}+1){\rm Ci}(r_-L) $$ $$+i(e^{-ir_-L}-1)({\rm Si}(r_-L)-\pi)]
+2i\pi(e^{-ir_-L}-1)\} $$
$$+{4\over{\epsilon}}g^4C_GTr(t_bt_d)n_{\beta}^*n_{\rho}{\pi}^
{2-{{\epsilon}\over 2}}
{(2\pi)}^{-n}{1\over{q_+r_-}}\times \{(e^{-iTq_+}-1)[(e^{-ir_-L}+1)
({\rm Ci}(p_-L)-{\rm Ci}(q_-L))$$ $$+i(e^{-ir_-L}-1)({\rm Si}(p_-L)-{\rm Si}(q_-L)
+\pi )] $$ $$ -(e^{-ir_-L}+1)[{\rm Ci}(r_-L)-{\rm Ci}(p_-L)+{\rm Ci}(q_-L)]
-i(e^{-ir_-L}-1)[{\rm Si}(r_-L)-{\rm Si}(p_-L)+{\rm Si}(q_-L)]\} $$
$$ -{4\over{\epsilon}}g^4C_GTr(t_bt_d)n_{\rho}^*n_{\beta}{\pi}^
{2-{{\epsilon}\over 2}}{(2\pi)}^{-n}{1\over{p_+r_-}}e^{-iTq_+} $$ $$
\times\{(e^{-iTp_+}-1)[(e^{-ir_-L}+1){\rm Ci}(r_-L)+i(e^{-ir_-L}-1)
({\rm Si}(r_-L)-\pi)] $$ $$ +(e^{-ir_-L}+1)[{\rm Ci}(p_-L)-{\rm Ci}(q_-L)
+{\rm Ci}(r_-L)]+i(e^{-ir_-L}-1)[{\rm Si}(p_-L)-{\rm Si}(q_-L)+{\rm Si}(r_-L)]
\} \eqno(B1) $$

\beginsection  Appendix C

In the case of Wilson loops with spacelike and/or timelike lines strict
application of dimensional regularization is not possible.
 As an example let us take the
self-energy type of graph in the triangle Wilson loop with one 
spacelike\footnote{$^*$}{This feature of the M-L prescription was noticed
already in A. Andra\v si, hep-th 9411117, unpublished} and
two lightlike sides shown in Fig.20.

$$ W_{\beta \rho}=C_{\beta \rho}\int d^nk{1\over{k^2+i\eta}}
{{k_-}\over{k_++i\omega k_-}}{1\over{k_3p_3}} $$ $$ \times \{{1\over{(p-k)_3}}
e^{-ip_3L}(e^{i(p-k)_3L}-1)+{1\over{(p+k)_3}}(e^{-i(p+k)_3L}-1)\} $$
$$ =C_{\beta \rho}W  \eqno(C1) $$
where
$$ C_{\beta \rho}=-ig^4v_{\beta}n_{\rho}Tr(t_bt_d){(2\pi)}^{-n}{1\over{q_+}}
(e^{{iq_+L}\over 2}-1)e^{-{{iq_-L}\over 2}}  \eqno(C2) $$
There are two poles in the upper-half complex $ k_0 $ plane.
\vskip 1cm
(a) $$ k^2+i\eta=0,$$ $$ k_0=-k+i\eta $$
\vskip 1cm
(b) $$ k_++i\omega k_-=0, $$ $$ k_0=-k_3+2i\omega k_3 \theta (k_3)
\eqno(C3) $$
Let us take the first part of $ W $ with the $ {1\over{(p-k)_3}} $ denominator.
After the $ k_0 $ integration it gives
$$ W_1=2i\pi e^{-ip_3L}\int dk_3d^{2-\epsilon}K{1\over{2k}}
{{k+k_3}\over{k-k_3+i\omega (k+k_3)}} $$ $$ \times {1\over{k_3p_3(p-k)_3}}
\{\cos (p-k)_3L-1+i\sin (p-k)_3L\} $$ $$ +2i\pi e^{-ip_3L}\int_{0}^{\infty}
dk_3\int d^{2-\epsilon}K{{2k_3}\over{K^2+4i\omega k_3^2-i\eta}} $$ $$
\times{1\over{k_3p_3(p-k)_3}}\{\cos (p-k)_3L-1+i\sin (p-k)_3L\} \eqno(C4) $$
Naively, one would strictly apply the rules of dimensional regularization and
set the second integral to zero as a tadpole in the perpendicular momentum
$ K $. However, after the introduction of polar coordinates
$$ k_3=k\cos \theta=kx, $$ $$ d^{3-\epsilon}k=k^{2-\epsilon}dk
(1-x^2)^{-{{\epsilon}\over 2}}dx \int d\Phi, $$ $$ \int d\Phi=
{{2{\pi}^{1-{{\epsilon}\over 2}}}\over{\Gamma (1-{{\epsilon}\over 2})}} 
 \eqno(C5) $$
and integration over $ k $, the first integral leads to an integral which is
not defined for any $ \epsilon $.
\vskip 0.5cm
 Therefore we have to keep $ \omega $ in the integrand and it becomes a part of
the gauge. We can choose two ways. Either we evaluate integrals separately
in terms of the spurious, 'ambiguous' terms of the form $ \omega^{-
{{\epsilon}\over 2}}{\epsilon}^{-2} $ dictated by the tadpole
$$ W^T=\int d^{2-\epsilon}K{1\over{K^2+4i\omega k_3^2-i\eta}}=
{\pi}^{1-{{\epsilon}\over 2}}\Gamma({{\epsilon}\over 2}){\omega}^
{-{{\epsilon}\over 2}}2^{-\epsilon}e^{-{i\pi \epsilon}\over 4}k_3^{-\epsilon},
\eqno(C6) $$
or we transform the tadpole into polar coordinates
$$ W^T=\int_{0}^{\infty}dk k^{-\epsilon}\int_{0}^{1}dx(1-x^2)^
{-{{\epsilon}\over 2}}{1\over{1-x^2+4i\omega x^2}}\int d\Phi \eqno(C7) $$
and sum it up with the first integral in (C4) leading to
$$ W_1=-i\pi e^{-ip_3L}p_3^{-1-\epsilon}[{\rm ci}(p_3L)+{1\over{\epsilon}}
+i{\rm si}(p_3L)+i\pi]\int d\Phi $$ $$ \times \int_{0}^{1} dx(1-x^2)^
{-{{\epsilon}\over 2}}x^{\epsilon -2}[{{2(1+x^2)}\over{1-x^2+2i\omega (1+x^2)}}
-{{4x}\over{1-x^2+4i\omega x^2}}] $$ $$+{(i\pi)}^2 e^{-ip_3L}p_3^{-1-\epsilon}
\int_{0}^{1} dx(1-x^2)^{-{{\epsilon}\over 2}}x^{\epsilon -2}
{{1-x}\over{1+x+i\omega (1-x)}}\int d\Phi . \eqno(C8) $$
\vskip 0.5cm
We notice how crucial the contribution from the tadpole
$ -{{4x}\over{1-x^2+4i\omega x^2}} $ is for the regularization of the pole
at $ x=1 $. Only after the addition of the tadpole we can set $ \omega =0 $
in (C8) and evaluate the integrals in the strip $ 1<\epsilon <4 $.
In the same way we evaluate the second part of (C1) with $ {1\over{(p+k)_3}} $
denominator.\footnote{$^*$}{Let us mention that the change of variable
$ k_3$ into $-k_3 $ in the second part of (C1) is not permitted as it creates the
pole in $ k_3 $.}
\vskip 0.5cm
Thus we obtain the result for (C1)
$$ W_{\beta \rho}=-C_{\beta \rho}{{4i{\pi}^{2-{{\epsilon}\over 2}}}\over
{\Gamma(1-{{\epsilon}\over 2})}}p_3^{-1-\epsilon}2^{-\epsilon}[
{2\over{\epsilon}}+1] $$ $$ \times \{(e^{-ip_3L}-1)({\rm ci}(p_3L)+
{1\over{\epsilon}})+i(e^{-ip_3L}+1)({\rm si}(p_3L)+{{\pi}\over 2})\}.
\eqno(C9) $$
This graph in the Feynman gauge contains only simple single poles. Hence,
the funny non-local sine and cosine divergences and the double pole are
caused by the choice of the M-L gauge, not the lightlike sides of the loop.

\beginsection  References

[1] A. Andra\v si and J.C. Taylor, Nucl. Phys.{\bf B516}(1998)417;

\noindent
[2] V.S. Dotsenko and S.N. Vergeles, Nucl. Phys. {\bf B169}(1980)527;

\noindent
[3] R. Brandt, F. Neri and Masa-aki Sato, Phys. Rev.{\bf D24}(1981)879;

\noindent
[4] G. Leibbrandt, Phys. Rev. {\bf D29}(1984)699;

\noindent
[5] J.G.M. Gatheral, Phys. Lett.{\bf 133B}(1983)90;

\noindent
[6] J. Frenkel and J.C. Taylor, Nucl. Phys.{\bf B246}(1984)231;

\noindent
[7] S. Mandelstam, Nucl. Phys.{\bf B213}(1983)149;   

\noindent
[8] A. Andra\v si, G. Leibbrandt and S.L. Nyeo, Nucl. Phys.{\bf B276}
(1986)445;

\noindent
[9] A. Andra\v si and J.C. Taylor, Nucl. Phys.{\bf B302}(1988)123;
 
\noindent
[10] P. Gaigg, W. Kummer and M. Schweda (Eds.), 

\line{\hskip 0.5 cm Physical and Nonstandard Gauges, Proceedings, Vienna,
Austria 1989 \hfill}

\vskip 1cm

\beginsection Figure Captions

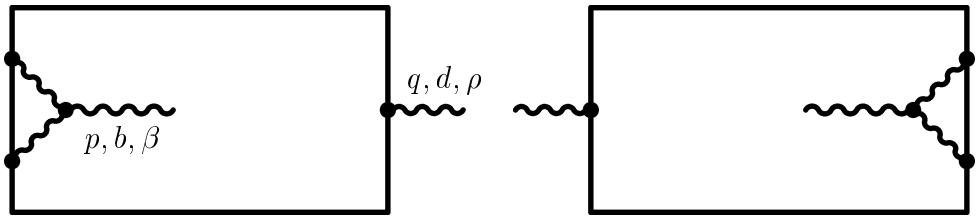
Fig.1. Wilson operator at order $ g^4 $ with two real gluons and one
3-gluon vertex. The sides of the loop are along the lightlike vectors used
to define the M-L prescription, $ n^* $ of length $ L $ and $n $ of
length $ T $. The two graphs of the $A$ -set which contribute to the 
$ n^*n^* $ sector have their symmetric counterparts.

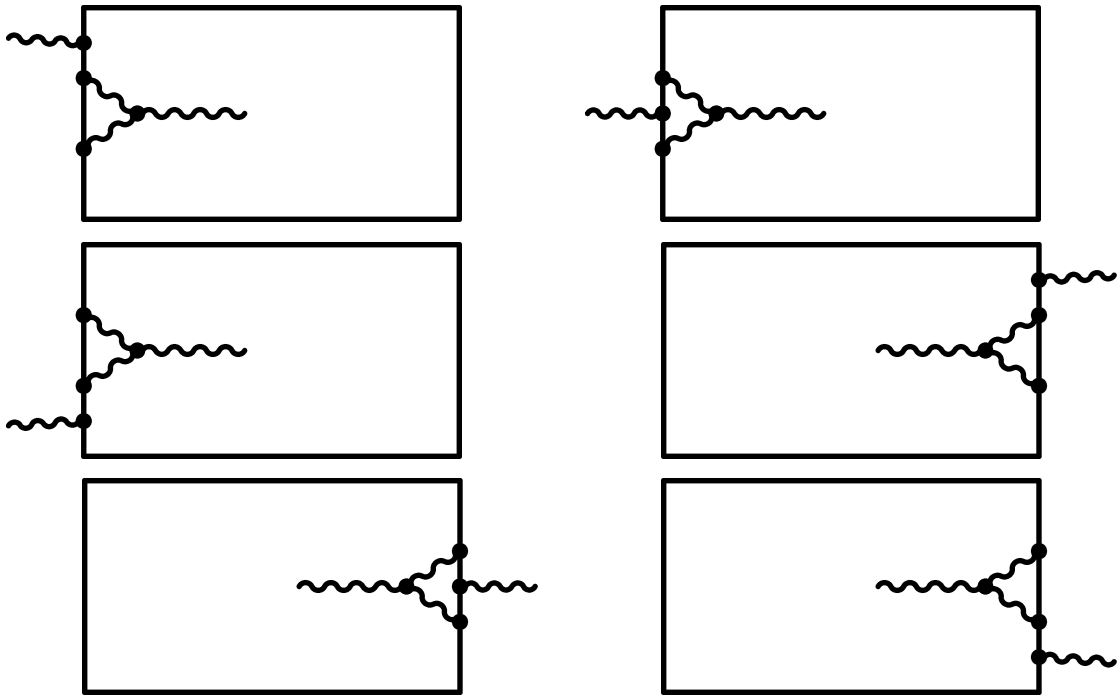
Fig.2. $ B$-set of graphs.

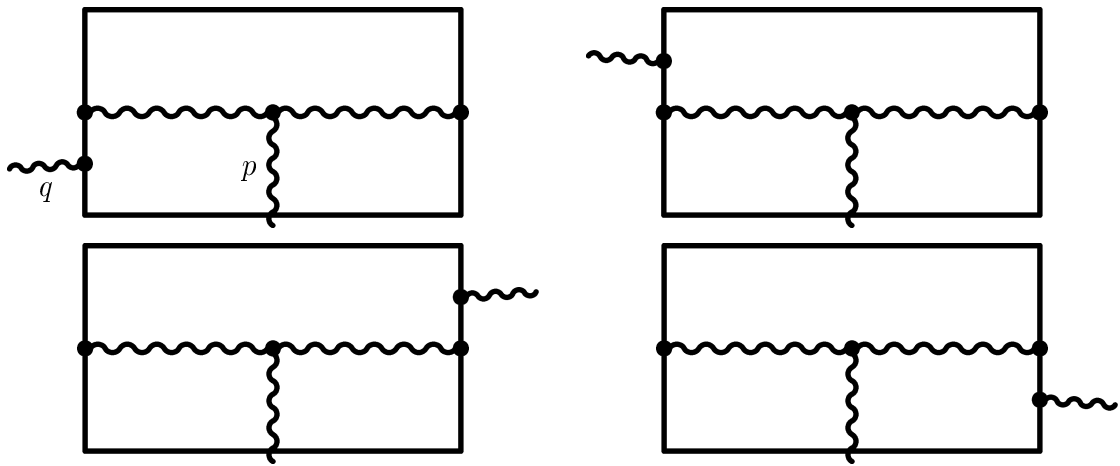
Fig.3. $ C $-set.

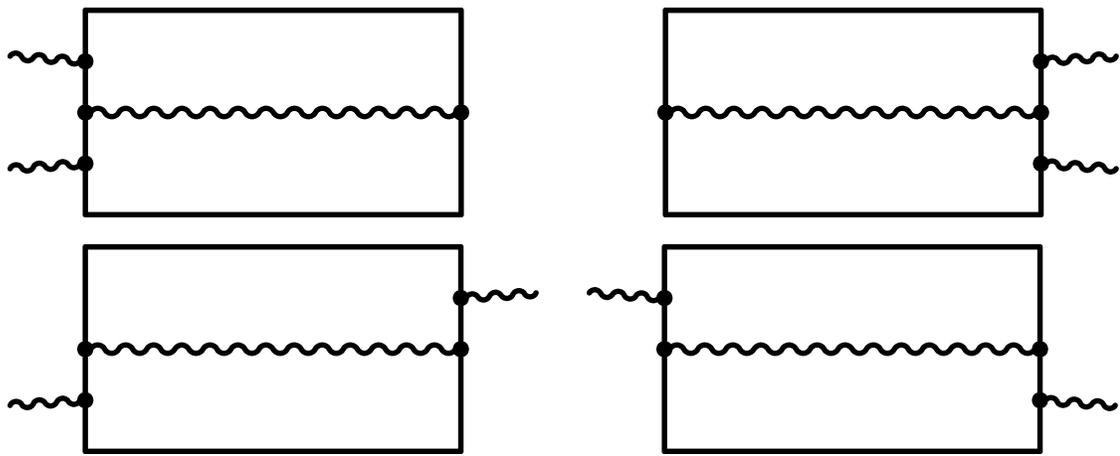
Fig.4. $ D $-set

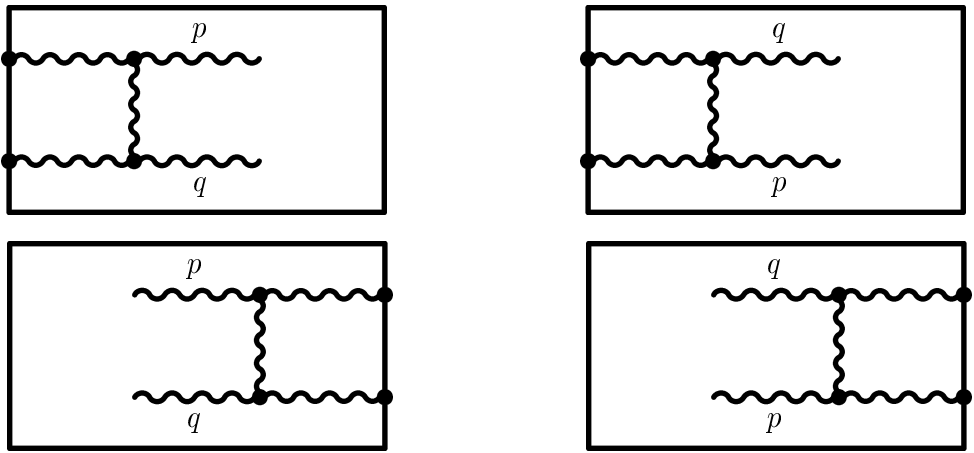
Fig.5. $ E $-set

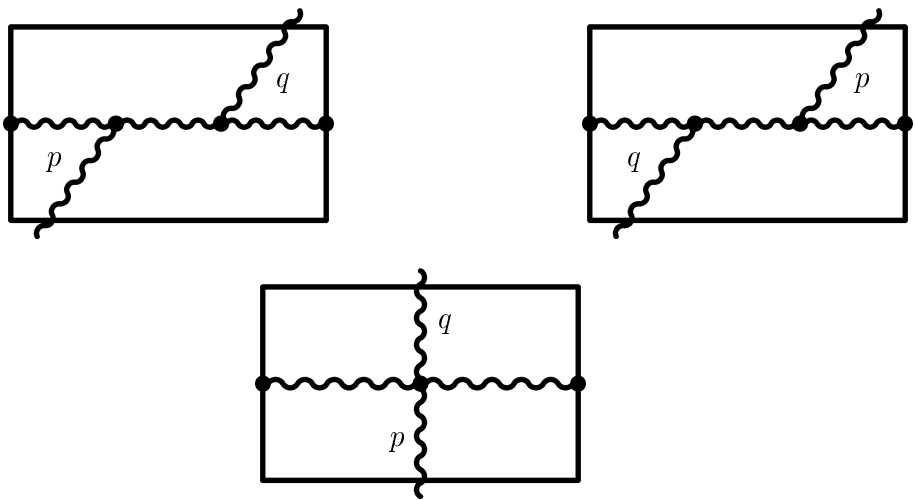
Fig.6. Graphs with two 3-gluon vertices and a graph with the 4-gluon vertex
which contribute to the $ nn $ sector of the Wilson operator at order $g^4$.

Figs.7 and 8. $ G1 $-set of graphs. Graphs with one 3-gluon vertex which
contribute to the $ nn $ sector.

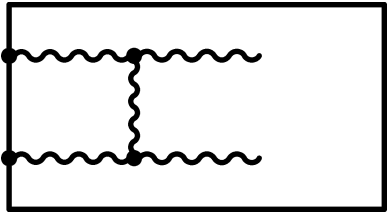
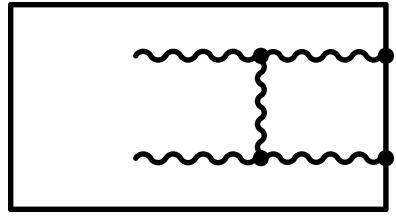
Fig.9. Left and right graphs with two 3-gluon vertices in the $ G(L+R) $ -set
which contribute to the $ nn $ sector.

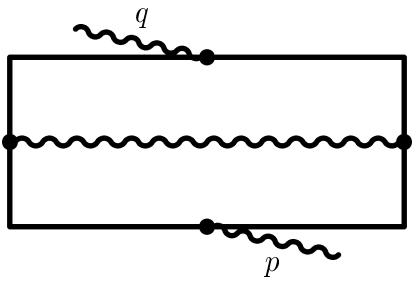
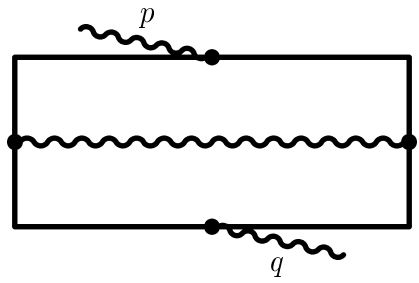
Fig.10. $ G0 $-set of graphs in the $ nn $ sector.

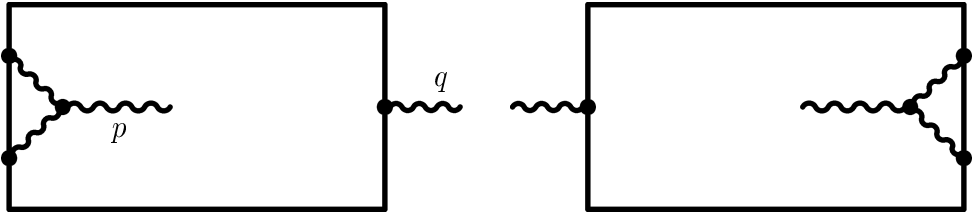
Fig.11. The $ A $-set of graphs which contribute to the 
$n_{\beta}n^*_{\rho}$ sector.

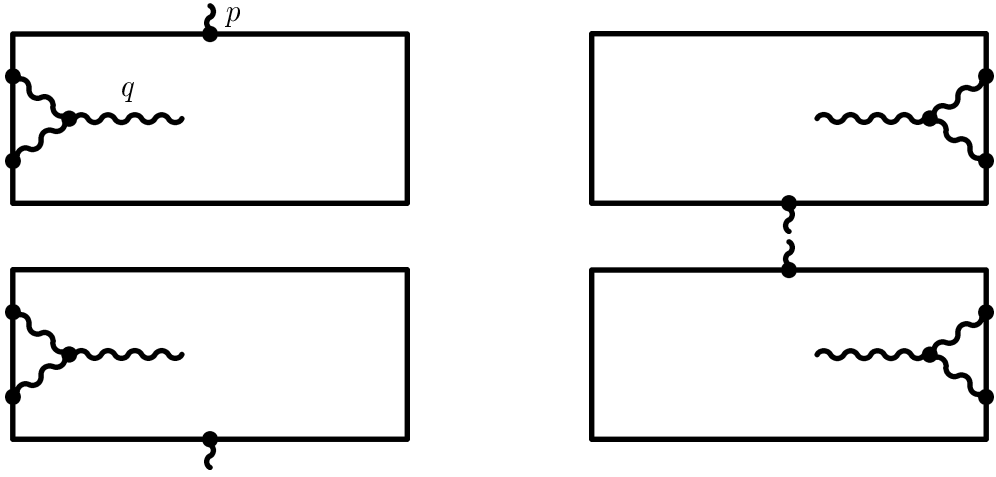
Fig.12. The $ A' $-set of graphs.

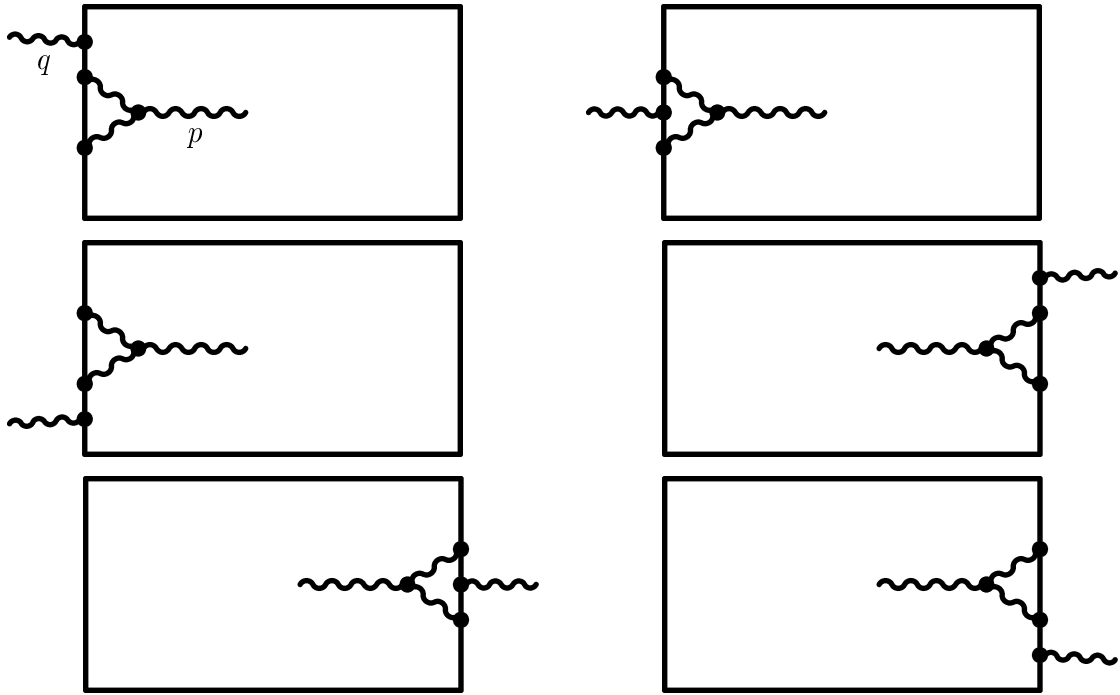
Fig.13. $ B $-set of graphs in the $ n_{\beta}n^*_{\rho} $ sector.

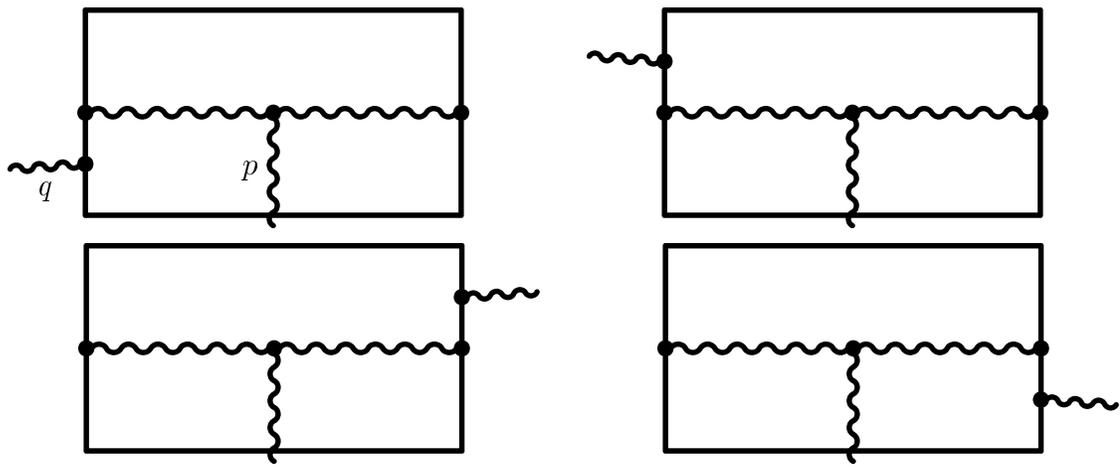
Fig.14. $ C $-set of graphs.

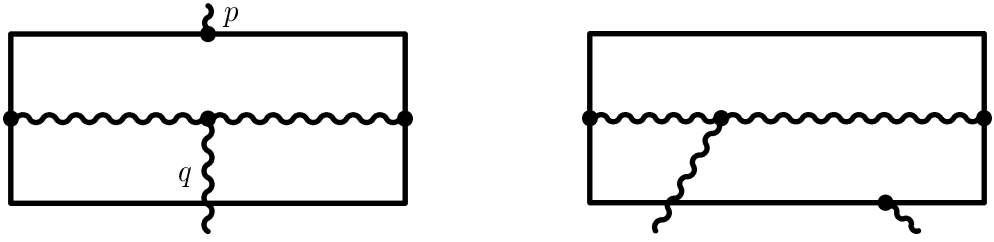
Fig.15. $ C' $-set.

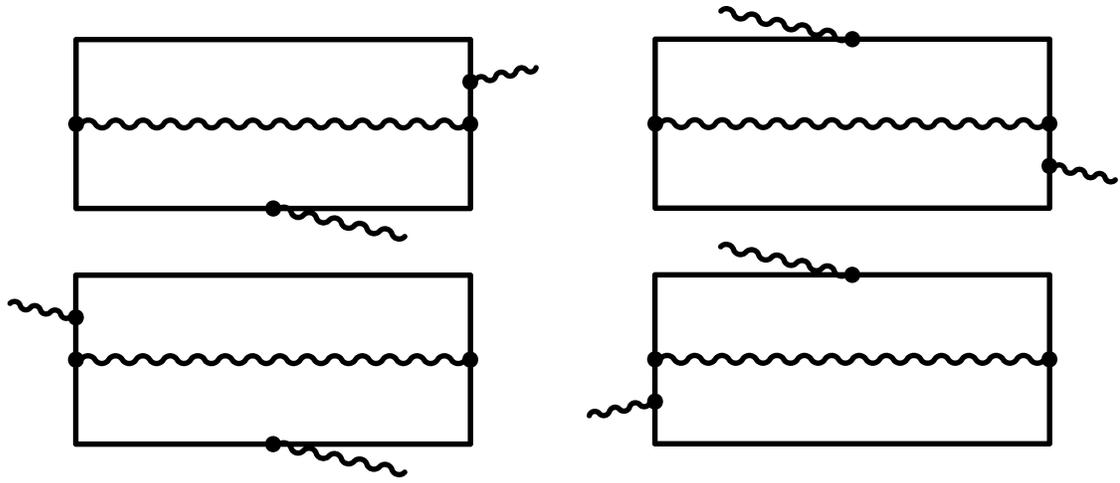
Fig.16. $ D $-set.

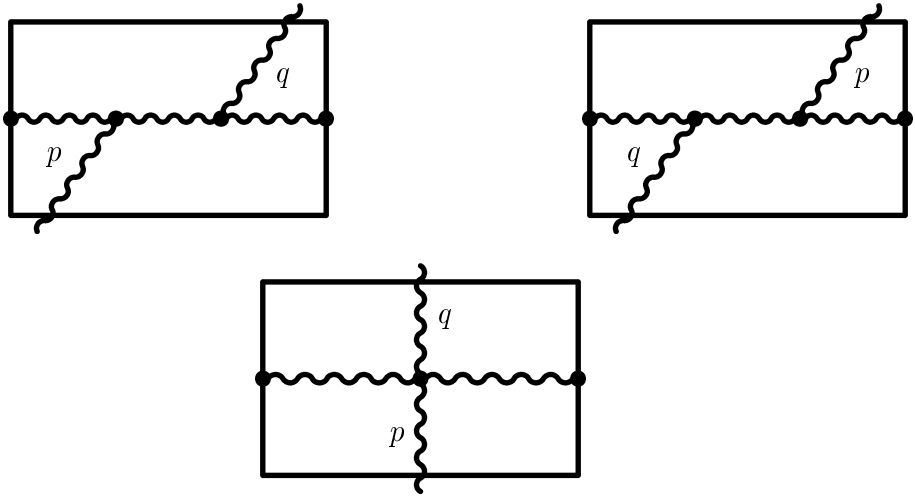
Fig.17. $ E $-set.

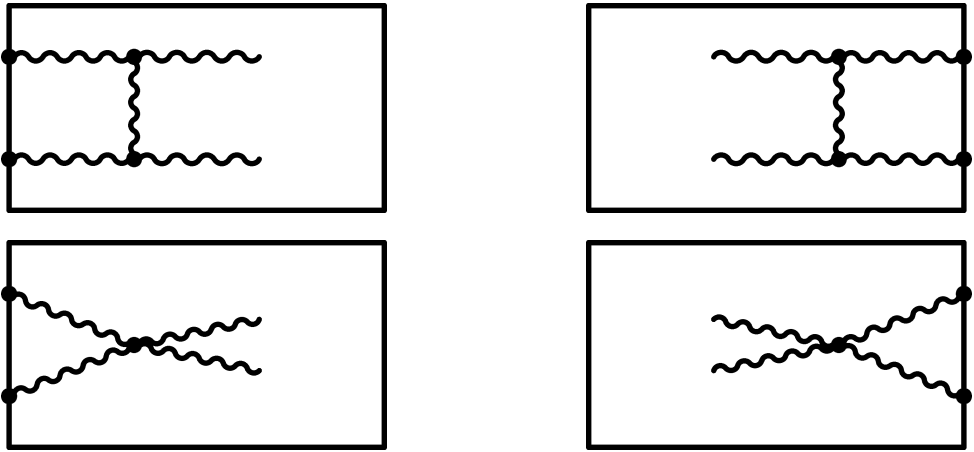
Fig.18. $ F $-set of graphs which contribute to the $ n_{\beta}n^*_{\rho} $
sector.

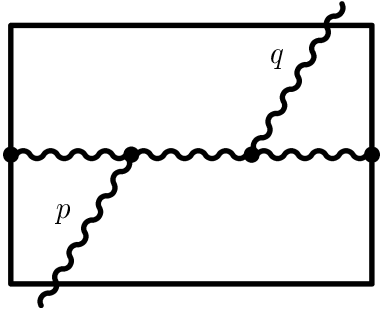
Fig.19. The graph with two 3-gluon vertices which in the Feynman gauge contains
{\bf no} UV divergences. The same graph in the M-L gauge contains double
divergences and single divergences with non-local $ {\rm Ci} $ and ${\rm Si}$
functions.

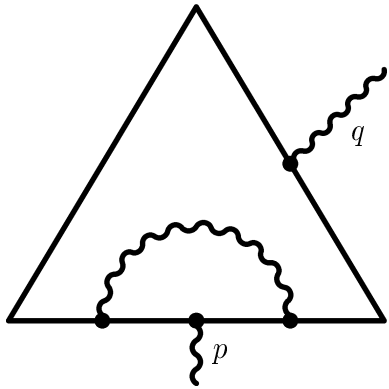
Fig.20. The triangle Wilson operator. The base is along the spacelike vector
$v_{\beta}$ of length $L$, while the sides are along the lightlike vectors
$n$ and $n^*$ of length ${L\over 2}$.

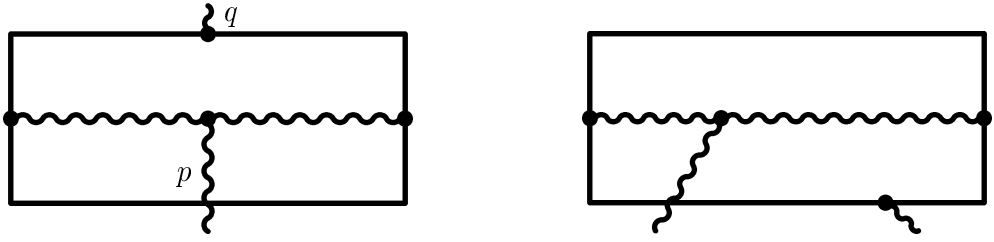
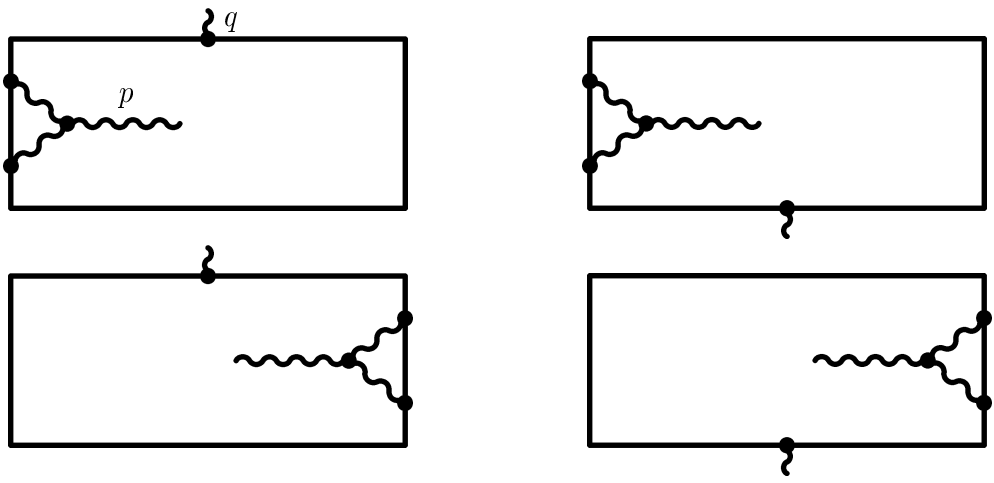
\bye